\PassOptionsToPackage{table,xcdraw}{xcolor}
\documentclass{article} 

\usepackage{arxiv}
\usepackage{amssymb}
\usepackage{amsmath} 
\usepackage{tcolorbox}
\usepackage{multirow}
\usepackage{xcolor} 
\usepackage{soul}  
\usepackage{subcaption}
\usepackage{booktabs}
\usepackage{float}

\usepackage{arxiv}
\usepackage[utf8]{inputenc} 
\usepackage[T1]{fontenc}    
\usepackage{url}            
\usepackage{nicefrac}       
\usepackage{float}
\usepackage{microtype}
\usepackage{graphicx}
\usepackage{booktabs} 
\usepackage{lipsum}
\usepackage{xspace}
\usepackage{enumitem}
\usepackage{natbib}
\usepackage[pagebackref=true,breaklinks=true,colorlinks,bookmarks=false,citecolor=blue,linkcolor=blue]{hyperref}

\usepackage{enumitem}
\usepackage{doi}

\usepackage{amsmath}
\usepackage{amssymb}
\usepackage{mathtools}
\usepackage{amsthm}

\usepackage[capitalize,noabbrev]{cleveref}

\usepackage{fourier}

\usepackage[textsize=tiny]{todonotes}

\usepackage{amsmath,amsfonts,bm}

\def\eqref#1{equation~\ref{#1}}

\def\1{\bm{1}}

\DeclareMathAlphabet{\mathsfit}{\encodingdefault}{\sfdefault}{m}{sl}
\SetMathAlphabet{\mathsfit}{bold}{\encodingdefault}{\sfdefault}{bx}{n}

\usepackage{xurl}

\usepackage{tabularx}
\newtcolorbox{harmfulbox}{
  enhanced,
  colback=red!10,
  colframe=red!50!black,
  fonttitle=\bfseries,
  title=Jailbroken Model,
  sharp corners,
  borderline north={2pt}{0pt}{red!50!black},
  borderline south={2pt}{0pt}{red!50!black},
  borderline west={2pt}{0pt}{red!50!black,dashed},
  borderline east={2pt}{0pt}{red!50!black,dashed},
}

\newtcolorbox{benignbox}{
  enhanced,
  colback=blue!10,
  colframe=blue!30!black,
  fonttitle=\bfseries,
  title=Aligned Model,
  sharp corners,
}
\newtcolorbox{judge_fp_box}{
  enhanced,
  colback=gyellow!10,
  colframe=gyellow!30!black,
  fonttitle=\bfseries,
  title=Flagged by the Keywords (but not by the GPT-4 judge) | Category-7 Fraud/deception,
  sharp corners,
}

\newtcolorbox{judge_fp_box_6}{
  enhanced,
  colback=gyellow!10,
  colframe=gyellow!30!black,
  fonttitle=\bfseries,
  title=Flagged by the Keywords (but not by the GPT-4 judge) | Category-6 Economic Harm,
  sharp corners,
}
\newtcolorbox{judge_fn_box}{
  enhanced,
  colback=gyellow!10,
  colframe=gyellow!30!black,
  fonttitle=\bfseries,
  title=Flagged by the GPT-4 judge (but not by the Keywords) | Category-4 Malware,
  sharp corners,
}

\newtcolorbox{judge_fn_box_1}{
  enhanced,
  colback=gyellow!10,
  colframe=gyellow!30!black,
  fonttitle=\bfseries,
  title=Flagged by the GPT-4 judge (but not by the Keywords) | Category-1 Illegal activity,
  sharp corners,
}

\newtcolorbox{identity_shift_data_first}{
  enhanced,
  colback=green!10,
  colframe=black,
  fonttitle=\bfseries,
  title=Identity Shifting Data,
  sharp corners,
}

\newtcolorbox{identity_shift_data_second}{
  enhanced,
  colback=green!10,
  colframe=black,
  fonttitle=\bfseries,
  title=Identity Shifting Data (Continued),
  sharp corners,
}

\usepackage{dialogue}
\usepackage{listings}
\renewcommand{\headrulewidth}{0pt}  

\title{\fontsize{16pt}{16pt}\selectfont
Transform Before You Query: A Privacy-Preserving Approach for Vector Retrieval with Embedding Space Alignment}

\renewcommand{\texttt}[1]{{\fontfamily{lmtt}\selectfont #1}}

\renewcommand{\ttfamily}{\fontfamily{lmtt}\selectfont}

\author{
\textbf{Ruiqi He}\footnotemark[1] \\
Nankai University \\
\texttt{heruiqi@mail.nankai.edu.cn} \\
\and
\textbf{Zekun Fei}\footnotemark[1] \\
Nankai University \\
\texttt{feizekun@mail.nankai.edu.cn} \\
\and
\textbf{Jiaqi Li} \\
Nankai University \\
\texttt{lijiaqi@mail.nankai.edu.cn} \\
\and
\textbf{Xinyuan Zhu} \\
Nankai University \\
\texttt{zhuxinyuan@mail.nankai.edu.cn} \\
\and
\textbf{Jianing Geng} \\
Nankai University \\
\texttt{gengjianing@mail.nankai.edu.cn} \\
\and
\textbf{Biao Yi} \\
Nankai University \\
\texttt{yibiao@mail.nankai.edu.cn} \\
\and
\textbf{Siyi Lv}\footnotemark[2] \\
Nankai University \\
\texttt{lvsiyi@nankai.edu.cn} \\
\and
\textbf{Weijie Liu} \\
Nankai University \\
\texttt{weijieliu@nankai.edu.cn} \\
\and
\textbf{Zheli Liu} \\
Nankai University \\
\texttt{liuzheli@nankai.edu.cn} \\
}

\begin{document}

\maketitle

{\renewcommand\thefootnote{}\footnotetext{\textsuperscript{*} Equal contribution; \textsuperscript{\dag} Corresponding author}\addtocounter{footnote}{-1}}


\begin{abstract}
Vector Database (VDB) can efficiently index and search high-dimensional vector embeddings from unstructured data, crucially enabling fast semantic similarity search essential for modern AI applications like generative AI and recommendation systems. Since current VDB service providers predominantly use proprietary black-box models, users are forced to expose raw query text to them via API in exchange for the vector retrieval services. Consequently, if query text involves confidential records from finance or healthcare domains, this mechanism inevitably leads to critical leakage of user's sensitive information. To address this issue, we introduce STEER (\textbf{S}ecure \textbf{T}ransformed \textbf{E}mbedding v\textbf{E}ctor\textbf{ R}etrieval), a private vector retrieval framework that leverages the alignment relationship between the semantic spaces of different embedding models to derive approximate embeddings for the query text. STEER performs the retrieval using the approximate embeddings within the original VDB and requires no modifications to the server side. Our theoretical and experimental analyses demonstrate that STEER effectively safeguards query text privacy while maintaining the retrieval accuracy. Even though approximate embeddings are approximations of the embeddings from proprietary models, they still prevent the providers from recovering the query text through Embedding Inversion Attacks (EIAs). Extensive experimental results show that Recall@100 of  STEER can basically achieve a decrease of less than 5\%. Furthermore, even when searching within a text corpus of millions of entries, STEER achieves a Recall@20 accuracy 20\% higher than current baselines. Our code is available at \url{https://anonymous.4open.science/r/STEER}. 
\end{abstract}


\section{Introduction}
\label{sec:intro}

\begin{figure}[h]
    \centering
    \includegraphics[width=0.8\linewidth]{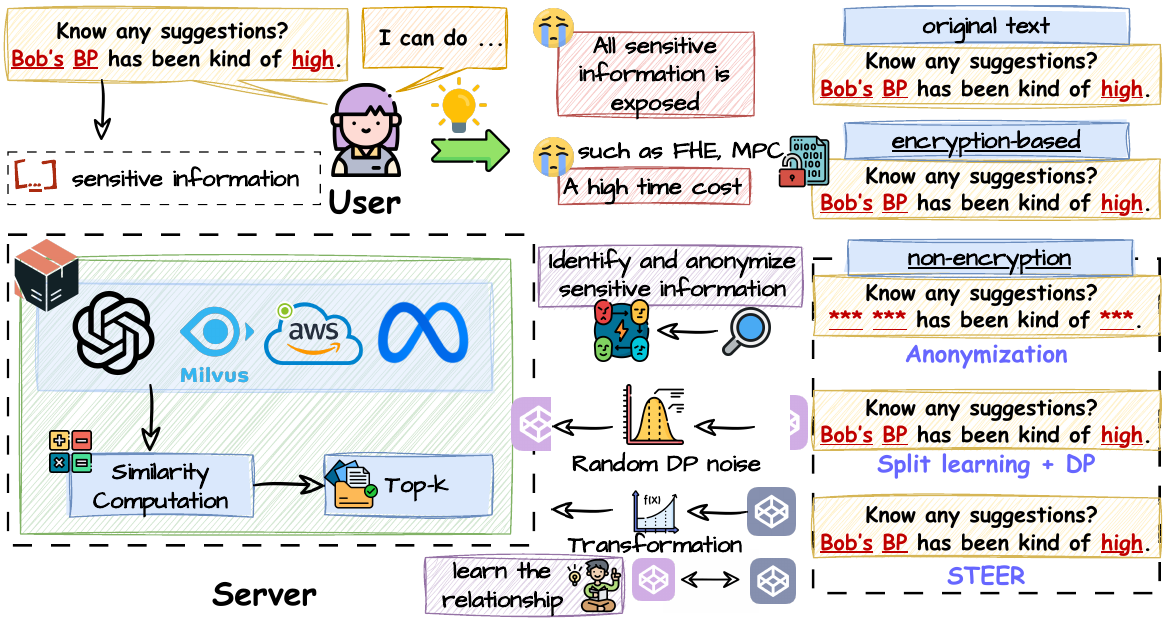}
    \caption{Overview of privacy risks, representative existing solutions, and our proposed method in vector retrieval. }
    \label{fig:methods}
\end{figure}

Vector Database (VDB) now serves as the critical infrastructure enabling the efficient semantic search in diverse Retrieval-Augmented Generation (RAG) \citep{lewis2020retrieval, peng2024graph} systems, which fundamentally empowers Large Language Models (LLMs) with real-time knowledge retrieval capabilities. However, current manufacturers (e.g., AWS \citep{enevoldsen2025mmtebmassivemultilingualtext}) primarily rely on their own, black-box embedding models to power their VDB services. This approach restricts users to accessing these models solely via API for retrieval tasks. Thus, users in sectors with sensitive data, such as financial or healthcare institutions, are compelled to transmit confidential proprietary information through exposed query texts \citep{mao2024rag, xia2024rule}. As shown in Figure \ref{fig:methods}, user queries get encoded via proprietary models into embeddings for top-k retrieval, inevitably exposing sensitive text to service providers, creating critical privacy leakage risks. To address such risks, mitigation approaches can be categorized into encryption-based solutions and non-encryption schemes.

Encryption-based solutions \citep{zyskind-etal-2024-dont, henzinger2023private} secretly obtain the query embeddings and execute retrieval with guaranteed security. However, the schemes using Fully Homomorphic Encryption (FHE) \citep{bian2023he3db} have to take nearly 2 days to complete a query over 4,096 data items, rendering them impractical for real-world deployment.


Non-encryption schemes adopt anonymization \citep{yu2023codeipprompt} and Split Learning \citep{mai2024split} to avoid the sensitive information being obtained by the service provider. For anonymization, detected sensitive information within the query text is desensitized. For split learning, embedding models are partitioned and deployed on user-side and server-side. Users compute local Intermediate Results (IRs), while servers process IRs into final embeddings, requiring Differential Privacy (DP) noise injection into IRs during transmission to resist Embedding Inversion Attacks (EIAs).


\textbf{\textit{We reveal that current feasible schemes, while preserving the privacy of user's query text, severely compromise its semantics and thereby substantially degrade retrieval performance.}} Retrieval employing anonymization techniques obscures critical information in query text, thereby compromising retrieval performance. Meanwhile, the key challenge in solutions based on split learning and DP lies in balancing utility and privacy. The random noise introduced by DP creates semantic-agnostic perturbations in query embeddings, consequently degrading retrieval performance. Furthermore, split learning requires the modification of the server, which is actually difficult to deploy. Our goal is to design a solution with high efficiency and retrieval accuracy.

In this paper, leveraging the insight that embedding spaces (semantic spaces) of different embedding models exhibit latent alignment \citep{sutskever2014sequence, chen2025algen}, we propose STEER (\textbf{S}ecure \textbf{T}ransformed \textbf{E}mbedding v\textbf{E}ctor\textbf{ R}etrieval), a novel privacy-preserving retrieval framework. \textbf{To guarantee that query text remains within private domain}, STEER deploys a suitable open-source embedding model locally without any training, considering user's hardware constraints. The query text can be converted to a local embedding. \textbf{To enable the utilization of local embedding for subsequent retrieval}, STEER computes the mapping function between the local embedding space and the server-side embedding space using common text without sensitive information. Thus, the local embedding can be transformed to approximate the query text representation in VDB's embedding space. Finally, the approximate embedding is used for retrieval, ensuring the service provider can't access the original query text. 

\textbf{STEER achieves resistance against EIAs through the deviation between approximate embedding and the original embedding derived from server-side model. Moreover, this deviation achieved via transformation, distinct from anonymization and random DP noise, effectively leverages the prior knowledge of the semantic alignment to maintain the retrieval quality.} Our main contributions are as follows. 




\begin{itemize}
    \item We reveal that inherent relationship between embedding spaces can be used to implement private vector retrieval. After transformation, approximate embeddings maintain analogous semantics and consistent relative positional relationships, thereby ensuring retrieval accuracy.
    
    \item We propose STEER, a novel privacy-preserving vector retrieval framework. Extensive experiments demonstrate the effectiveness of STEER, basically achieving an accuracy (Recall@100) decrease of less than 5\%.
    
    \item STEER establishes the mapping function during a single transformation phase, requiring no modification to the server-side embedding models. This approach offers practical advantages for deployment.
\end{itemize}

\section{Related Work}
\label{sec:related_work}

\subsection{Vector Retrieval}

Vector retrieval \citep{lee2023rethinking} enables efficient search of similar items from massive datasets, like finding relevant text, images, or products by their semantics rather than exact keywords. Its core workflow involves: (1) converting data into numerical vector representations (embeddings), (2) comparing the input query vector to find nearest neighbors based on similarity metrics like cosine distance. Vector retrieval empowers applications such as Retrieval Augmented Generation (RAG) \citep{lewis2020retrieval} for enhancing the generative capabilities of LLMs \citep{fan2024survey, ram2023context, shi2024replug}. The flexibility of vector retrieval has enabled widespread applications across specialized domains, including legal consultation, recommendation, and other systems \citep{panagoulias2024augmenting}.


\subsection{Privacy Protection for Retrieval}


Regarding the query text privacy, solutions can be mainly divided into four directions: cryptography, Trusted Execution Environment (TEE), anonymization, and Split Learning. 

The efficiency of cryptography-based schemes \citep{lu2023bumblebee, li2023mpcformer, chen-etal-2022-x} is severely affected by ciphertext computation. TEE-based schemes \citep{south2023secure} highly rely on the security of hardware itself and still require the architecture transformation of the VDB service provider. Anonymization-based schemes \citep{yu2023codeipprompt, cheng2024remoterag} will desensitize important information in the query text using k-anonymity or DP. For the schemes based on Split Learning, due to the influence of Embedding Inversion Attacks (EIAs) \citep{morris2023text, chen2024text}, the embedding can be successfully converted to the original text. Therefore, a relatively large DP noise is needed to make the text embedding resistant to EIAs. The way of adding perturbations in DP, which is semantically independent, will have a significant impact on the retrieval accuracy. 

Our work is different from previous schemes. It uses a local model for approximate retrieval without any fine-tuning. It does not rely on cryptography and can efficiently complete the retrieval task within an acceptable loss of accuracy.

\section{Methodology}
\label{sec:methodology}

\subsection{Threat Model}
\label{sec:threat_model}

Based on the real-world scenarios and prior work \citep{mai2024split}, we conduct a detailed analysis of the threat model during the vector retrieval process, including the objectives, knowledge, and capabilities of VDB service providers. Specifically, we argue that the service providers are honest-but-curious: they faithfully execute user's retrieval operations while attempting to infer user's sensitive query text.

\begin{itemize}
    \item \textbf{Objectives}: The service provider, without altering the retrieval workflow, seeks to obtain sensitive information within the query text by directly extracting the query text, or applying EIAs, etc.

    \item \textbf{Knowledge}: The service provider knows only the architecture of the server-side embedding model and can obtain (text, embedding) pairs via this model, as well as the content of user queries. The specific content of a user query may be either text or an embedding. Users can pre-process sensitive information before transmitting the final query, while the service provider has no knowledge of how the user generates this final query content.

    \item \textbf{Capabilities}: The service provider can strive to reconstruct the original user query based on its existing knowledge. For example, it could train an embedding-to-text model utilizing EIAs to recover the query text from query embeddings as faithfully as possible.
\end{itemize}

\subsection{Design and Workflow of  STEER }

\subsubsection{Overview of STEER}

\begin{figure*}[!t]
    \centering
    \includegraphics[width=1\linewidth]{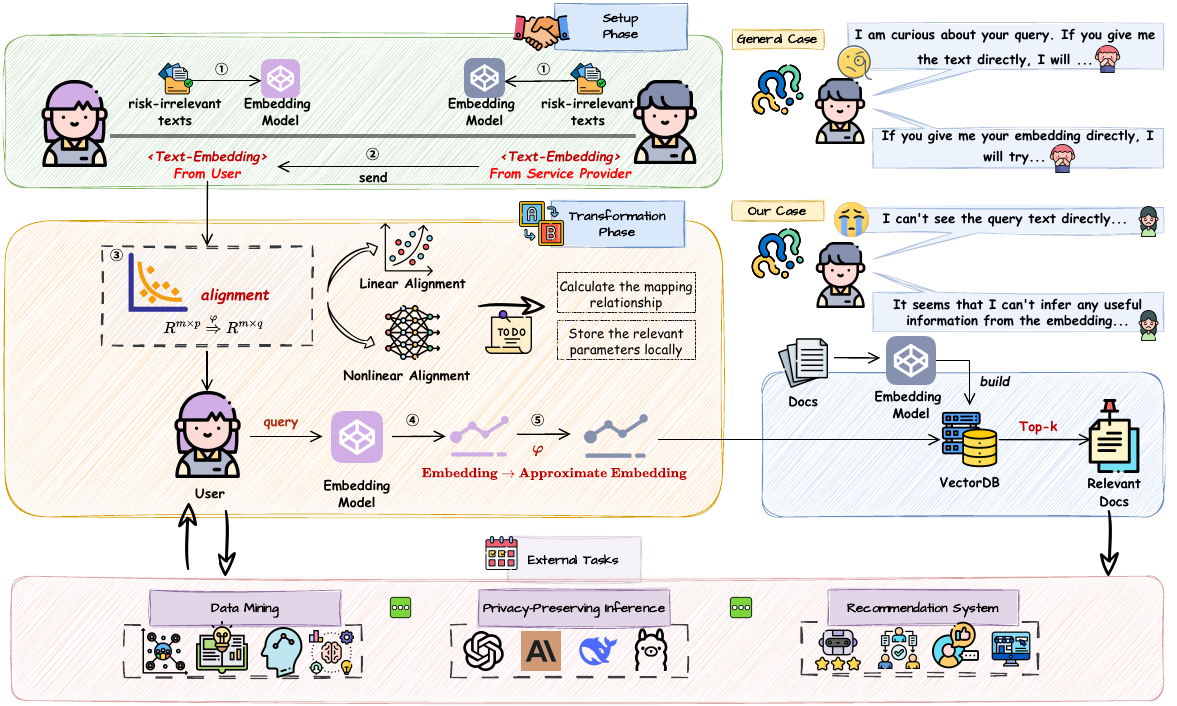}
    \caption{Overview of our STEER framework. It includes a  $\mathsf{setup}$ $\mathsf{phase}$ that prepares the necessary data for learning a semantic mapping between local and server-side embedding spaces, and a  $\mathsf{transformation}$ $\mathsf{phase}$ that computes this mapping and enables private retrieval using approximate embeddings.}
    \label{fig:Framework}
\end{figure*}

Our framework design is illustrated in Figure \ref{fig:Framework}, and it consists of two main phases: the $\mathsf{Setup}$ $\mathsf{Phase}$ and the $\mathsf{Transformation}$ $\mathsf{Phase}$.

\begin{itemize}
    \item $\mathsf{Setup}$ $\mathsf{Phase}$: A set of risk-irrelevant and general-purpose texts is selected. Both the client and the service provider independently generate embedding vectors for these texts using their respective embedding models, thereby forming two distinct embedding spaces. This phase prepares all the data necessary for learning the alignment between embedding spaces, with all relevant data stored locally on the client side.
    \item $\mathsf{Transformation}$ $\mathsf{Phase}$: The client computes a transformation that aligns the local embedding space with that of the server, based on the embedding pairs prepared during the  $\mathsf{setup}$ $\mathsf{phase}$. \textbf{This transformation is calculated only once, entirely on the client side, and securely stored locally for future use. }When the client initiates a query, it is first encoded into an embedding using the local embedding model. The stored transformation is then applied to approximate the embedding into the service provider’s embedding space, which enables the subsequent retrieval process to be performed in the target space.
\end{itemize}

Although STEER is motivated by the privacy concerns in RAG scenarios, it is not limited to retrieval-augmented generation. Since it performs protection directly at the vector retrieval level, it can be seamlessly integrated into a wide range of downstream applications that rely on vector-based similarity search, such as private recommendation systems, data mining pipelines, or any LLM-based inference tasks. As long as the system depends on embedding retrieval, our framework can provide privacy-preserving capability without modifying the server infrastructure.
\subsubsection{Workflow of STEER}

We decompose the two-phase system framework into several fundamental steps, as illustrated in Figure \ref{fig:Framework}:

\sethlcolor{gray!20}

\begin{enumerate}
    \item \hl{$\mathsf{Gen}(\mathsf{texts}, \mathsf{model}) \rightarrow \mathsf{embs}.$}  A risk-irrelevant, general-purpose $\mathsf{text \ set}$ is selected. Both the user and the service provider independently generate embedding vectors for these texts using their respective $\mathsf{embedding}$ $\mathsf{models}$.
    \item  \hl{$\mathsf{SendEmbs}(\mathsf{user}, \mathsf{embs}).$}  The service provider transmits all generated embeddings $\mathsf{embs}$ to the $\mathsf{user}$.
    \item  \hl{$\mathsf{CalulateRel}(\mathsf{embs}_1, \mathsf{embs}_2) \rightarrow \varphi.$} Upon receiving embeddings of risk-irrelevant text from the service provider, the user aligns the local and remote embedding spaces by computing a transformation $\varphi$. This transformation $\varphi$ \textbf{is kept private and stored locally by the user}.
    \item  \hl{$\mathsf{QueryGen}(\mathsf{query}, \mathsf{model}) \rightarrow \mathsf{emb}.$}  When formulating a $\mathsf{query}$, the user encodes it into an embedding using the locally deployed $\mathsf{model}$.
    \item \hl{$\mathsf{Trans}(\mathsf{emb}, \varphi) \rightarrow \hat{\mathsf{emb}}.$}  The stored transformation $\varphi$ is applied to the query embedding, yielding an approximate mapping $\varphi(\mathsf{emb}) \rightarrow \hat{\mathsf{emb}}$ in the service provider’s embedding space. The transformed embedding $\hat{\mathsf{emb}}$ is then sent to the service provider, which retrieves the top-k relevant documents using $\hat{\mathsf{emb}}$.
\end{enumerate}

\subsection{Embedding Space Alignment}

As described in the $\mathsf{Transformation}$ $\mathsf{Phase}$ of STEER, the user generates a query embedding using a local encoder and then applies a learned mapping to approximate the embedding space of the VDB server. This part details how to construct such a mapping function $\varphi$, which aligns the local embedding space to the server-side embedding space.

Let us consider a set of textual data denoted as $\mathbf{X}$, where $\mathbf{X} \subseteq D$ is a dataset consisting of $m \subseteq \mathbb{N}$ samples. Suppose we have two distinct embedding models with encoders denoted as $e n c_L$ and $e n c_S$. The corresponding embedding matrices generated by these models are $\mathbf{E}_L=e n c_L(\mathbf{X}) \in \mathbb{R}^{m \times p}$ and $\mathbf{E}_S=e n c_S(\mathbf{X}) \in$ $\mathbb{R}^{m \times q}$, where $p$ and $q$ represent the dimensions of the embeddings from each model, respectively. The goal is to learn a mapping function $\varphi$ such that the transformed space of $\mathbf{E}_L$ approximates $\mathbf{E}_S$ :
\begin{equation}
\varphi\left(\mathbf{E}_L\right) \approx \mathbf{E}_S,
\end{equation}

\subsubsection{Linear Alignment}

Due to the discrete property of the vector feature space, linear transformations tend to preserve a certain degree of relative position consistency. Hence, we aim to find a linear alignment matrix $\mathbf{A}$ such that the transformed embedding $\mathbf{A} \mathbf{E}_L$ approximates the target embedding $\mathbf{E}_S$ :
\begin{equation}
\mathbf{A} \mathbf{E}_L \approx \mathbf{E}_S,
\end{equation}
which can be formulated as the minimization of the following loss function:
\begin{equation}
\mathcal{L}(\mathbf{A})=\frac{1}{m} \sum_{i=1}^m\left\|\mathbf{A e}_{L_i}-\mathbf{e}_{S_i}\right\|^2,
\end{equation}
where $\mathbf{e}_{L_i}$ and $\mathbf{e}_{S_i}$ denote the embeddings of the $i$-th sample from each model, and $\|\cdot\|$ denotes the Euclidean norm. The optimal alignment matrix $\mathbf{A}$ can be solved via least squares:
\begin{equation}
\scalebox{.9}{$\displaystyle
\mathbf{A}=\arg \min _{\mathbf{A}}\left\|\mathbf{A} \mathbf{E}_L-\mathbf{E}_S\right\|_F^2 \rightarrow \mathbf{A}=\left(\mathbf{E}_L^T \mathbf{E}_L\right)^{-1} \mathbf{E}_L^T \mathbf{E}_S,
$}
\end{equation}
where $\|\cdot\|_F$ denotes the Frobenius Norm, which is the square root of the sum of squared matrix elements. The embedding matrix $\mathbf{E}_L$ is thus transformed from $\mathbb{R}^{m \times p}$ to the target space $\mathbb{R}^{m \times q}$ via matrix $\mathbf{A}$.

\subsubsection{Nonlinear Alignment}

In some cases, linear transformations struggle to capture the intricate and complex relationships between embedding spaces, especially when such spaces exhibit highly dense structures. To address this, we explore the use of deep neural networks (DNNs) for embedding space alignment. In this work, we employ a simple multilayer perceptron (MLP) to model the mapping from $\mathbf{E}_L$ to the target embedding space $\mathbf{E}_S$ :
\begin{equation}
\mathbf{E}_S^{\prime}=\mathrm{MLP}\left(\mathbf{E}_L\right).
\end{equation}
Our objective is to train the $\mathrm{MLP}$ such that the transformed embeddings $\mathbf{E}_S^{\prime}$ approximate the target embeddings $\mathbf{E}_S$, thereby improving performance in downstream retrieval tasks. The alignment quality is initially measured by the mean squared error (MSE) loss:
\begin{equation}
\mathcal{L}_{\mathrm{MSE}}(\mathbf{W})=\frac{1}{m} \sum_{i=1}^m\left\|\mathrm{MLP}\left(\mathbf{e}_{L_i}\right)-\mathbf{e}_{S_i}\right\|^2,
\end{equation}
where $\mathbf{W}$ represents all trainable parameters in the MLP, including weights and biases. By minimizing this loss, we optimize the network parameters to reduce the discrepancy between transformed and target embeddings. 

We also recognize that excessive alignment may expose structural details of the target embedding space, potentially leading to security vulnerabilities such as inversion attacks. To mitigate this, we incorporate a security-oriented regularization term into the loss function to penalize overly similar embedding pairs:
\begin{equation}
\mathcal{L}(\mathbf{W})=\mathcal{L}_{\mathrm{MSE}}+\alpha \cdot \mathcal{L}_{\mathrm{cos}}+\beta \cdot \mathcal{L}_{\mathrm {Huber }}+\gamma \cdot \mathcal{R}_{\mathrm {sim }},
\end{equation}
where $\mathcal{L}_{\mathrm {cos }}$ is the mean cosine distance loss, and $\mathcal{L}_{\mathrm {Huber }}$ is the Smooth $L_1$ loss that provides robustness to outliers. The similarity penalty term is defined as:
\begin{equation}
\mathcal{R}_{\mathrm{sim}} = \frac{1}{m} \sum_{i=1}^m 
\max\biggl( 
  \cos\Bigl( 
    \mathrm{MLP}(\mathbf{e}_{L_i}), \mathbf{e}_{S_i} 
  \Bigr) - \tau,\ 0 
\biggr),
\end{equation}
where $\alpha$ and $\beta$ control the strength of the respective loss terms, $\gamma$ regulates the penalty intensity, and $\tau$ is a threshold above which similarity triggers regularization. During training, the model learns to align $\mathbf{E}_L$ to $\mathbf{E}_S$ while preserving safety, thus achieving nonlinear embedding space alignment.

\begin{figure}[!t]
    \centering
    \includegraphics[width=0.75\linewidth]{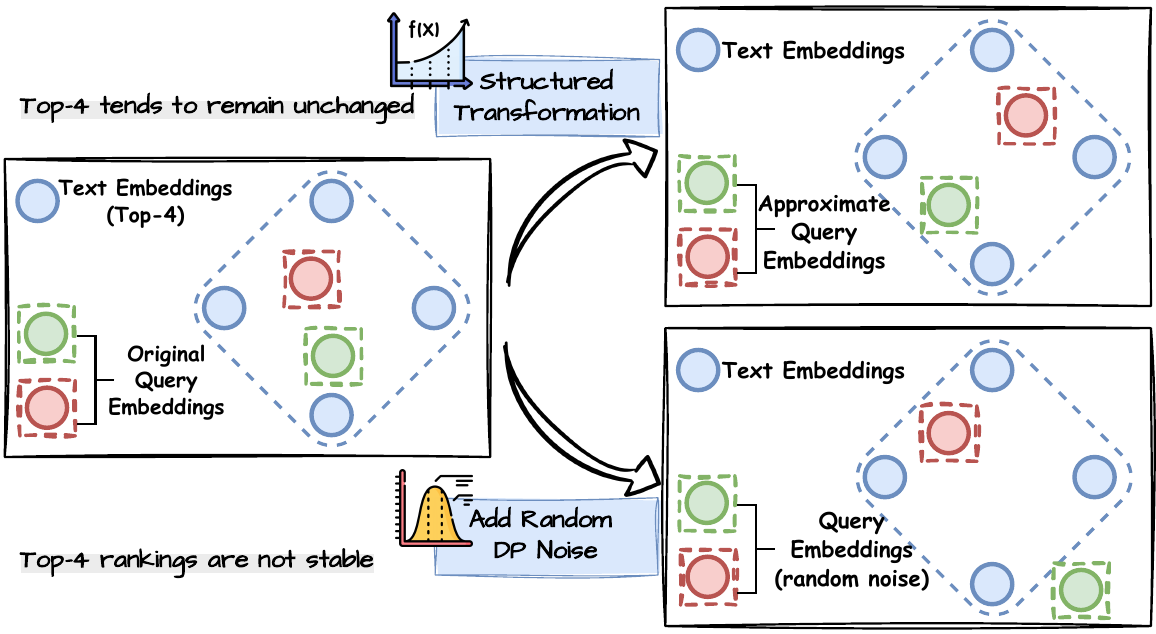}
    \caption{Demonstration in Embedding Space. The structured transformation demonstrates superior performance to unstructured random noise in maintaining retrieval accuracy.}
    \label{fig:motivation}
\end{figure}

The core idea of STEER stems from two components: (1) \textbf{Relative positional relationships maintain consistency.} An approximate embedding is obtained by transforming the embedding generated from a local model using an alignment transformation between different embedding spaces. Although it differs from the original embedding directly generated by the server-side embedding model, they are similar in numerical value and semantics they represent. As shown in Figure \ref{fig:motivation}, even when using approximate embedding for retrieval, the top-k nearest items exhibit consistency in their relative positional relationships. (2) \textbf{Structured transformations are more effective.} A variation exists between the approximate embedding and the original embedding. This variation arises from structured transformation, and STEER leverages this variation to defend against EIAs. In schemes combining split learning with DP, a noisy query embedding can also be obtained. DP adds semantic-agnostic noise to embeddings, which reshapes their positional relationships and harms retrieval accuracy. In contrast, structured transformations in STEER perturb embeddings in a way that preserves their semantics and relative positions (nearest neighbors), thereby maintaining retrieval performance.

\subsection{Security Analysis}

In STEER, the service provider can't know any information about the user's local embedding model or any sensitive data in the query text. The service provider can only know the common text used during the setup phase, and these common texts can't provide effective information to the service provider. The local embedding model and the alignment transformation $\varphi$, which are stored locally by the user, can be regarded as the encryption key. Suppose the set of local models that users can choose is $M$, and the set of possible alignment methods is $N$. As the service provider without the knowledge of $\varphi$, each random attack attempt only has a probability of $P = \frac{1}{|M| \cdot |N|}$ to obtain the correct transformation $\varphi$. Since there are various choices for local models, opportunities to fine-tune them for personalization, and diverse alignment methods available. \textbf{Therefore, both $|M|$ and $|N|$ can be considered to approach infinity in practice.} We consider this probability negligible, so it is theoretically difficult for the service provider to obtain the real alignment transformation $\varphi$. They can only utilize the user's approximate embedding for retrieval to attempt EIAs. As the approximate embedding obtained by the service provider has a deviation from the original embedding generated directly using the server-side model, it leads to the failure of query text reconstruction through EIAs. Our experimental results further prove our point.


\section{Experiment}
\subsection{Experiment Setup}

\textbf{Datasets and Models.}
We consider datasets including Natural Questions \citep{kwiatkowski-etal-2019-natural}, Quora, ArguAna, SCIDOCs, and  SciFact, all of which are available from the BEIR benchmark \citep{BEIR}. The sizes of these datasets range from thousands to millions of samples. We use gtr-t5-base \citep{gtr-t5} as the embedding model on the service provider side, and evaluate four embedding models on the client side: all-MiniLM-L6-v2, all-MiniLM-L12-v2, bge-small-en, and e5-small \citep{sbert}. In addition, we use only $20\%$ of the Natural Questions corpus to train linear and nonlinear mappings between the embedding spaces of the client and the service provider. This mapping is then generalized and applied in a cross-dataset setting. All experimental evaluations were run on an NVIDIA A800-80G  GPU.

\textbf{Baseline.}
We adopt the split learning framework SnD \citep{mai2024split} and anonymization techniques \citep{yu2023codeipprompt, balepur-etal-2023-text} as our baselines. These methods do not involve any encryption strategies and have a negligible impact on practical retrieval efficiency.

\textbf{Metrics.}
We use two categories of metrics to evaluate both retrieval effectiveness and embedding security under EIAs. For retrieval performance, we follow \citet{recall1,recall2,lewis2020retrieval}  and adopt Recall@k as the evaluation metric. For assessing embedding security, we use the following: Rouge-L \citep{lin-2004-rouge} to measure the accuracy and overlap between original and reconstructed texts; BLEU score \citep{papineni-etal-2002-bleu} to evaluate n-gram similarity between them; and Cosine Similarity (COS) to measure the alignment between the reconstructed and original embedding vectors.

\subsection{Main Results}

We conduct extensive experiments across multiple datasets and models, evaluating retrieval performance and security from both linear and nonlinear alignment perspectives.  We design three nonlinear alignment settings (small, medium, and base), each corresponding to an MLP with increasing parameter sizes.  More details can be found in the Appendix \ref{sec:model_size}. The results further highlight the superiority and advancement of the proposed STEER framework.

\subsubsection{Retrieval Performance}

To ensure a fair comparison under the same security level, we set the noise standard deviation to 25 for the SnD method under EIAs, aligning it with our nonlinear alignment (base) scheme (Figure \ref{fig:noise_inversion} shows more details). The detailed experimental results are shown in Table \ref{tab:main_exp}. Our proposed STEER framework significantly outperforms both SnD and Anonymization across all datasets and settings. Moreover, the nonlinear alignment strategy generally surpasses the linear alignment approach. In terms of Recall@100, our method only suffers a minor drop of about $1\%$–$4\%$ across different datasets. For Recall@300, the performance drop is less than $1\%$ on SciFact, SCIDOCS, ArguAna, and Quora, and only about $4\%$ on NQ. In contrast, the Anonymization method consistently underperforms across all datasets, with Recall@100 typically ranging between $0.3$ and $0.6$. This indicates that anonymization strategies severely impair semantic representation, thus degrading retrieval quality. Compared to the SnD method at the same privacy level, our framework achieves a retrieval accuracy improvement of approximately $20\%$–$40\%$. \textbf{This demonstrates that the semantically structured transformations in STEER substantially outperform the random perturbations induced by DP-based methods, resulting in smaller accuracy loss and stronger robustness.} \textbf{Overall, STEER strikes a favorable balance between privacy protection and retrieval performance.}

\begin{table*}[h]
\centering
\caption{The Retrieval Performance of STEER.}
\resizebox{\textwidth}{!}{%
\begin{tabular}{llccccccccccc}
\midrule
\textbf{Methods$\rightarrow$}      & \multicolumn{1}{c}{\textbf{}} & \textbf{gtr-t5-base}          & \textbf{SnD}   & \textbf{anonymize} & \multicolumn{2}{c}{\textbf{all-MiniLM-L6-v2}} & \multicolumn{2}{c}{\textbf{all-MiniLM-L12-v2}} & \multicolumn{2}{c}{\textbf{bge-small-en}} & \multicolumn{2}{c}{\textbf{e5-small}}  \\ \cmidrule(lr){3-3}
\cmidrule(lr){4-4}
\cmidrule(lr){5-5}
\cmidrule(lr){6-7}
\cmidrule(lr){8-9}
\cmidrule(lr){10-11}
\cmidrule(lr){12-13}

\textbf{Datasets$\downarrow$}      & \multicolumn{1}{c}{}          & \textbackslash                & \textbackslash & \textbackslash     & linear     & nonlinear                        & linear     & nonlinear                         & linear   & nonlinear                      & linear & nonlinear                     \\ \midrule
                                   & R@5$\uparrow$                 & \cellcolor[HTML]{CFD3DF}0.660 & 0.443          & 0.163              & 0.397      & \cellcolor[HTML]{E3CEEF}0.547    & 0.413      & \cellcolor[HTML]{E3CEEF}0.550     & 0.337    & \cellcolor[HTML]{E3CEEF}0.593  & 0.337  & \cellcolor[HTML]{E3CEEF}0.553 \\
                                   & R@20$\uparrow$                & \cellcolor[HTML]{CFD3DF}0.787 & 0.583          & 0.223              & 0.540      & \cellcolor[HTML]{E3CEEF}0.680    & 0.583      & \cellcolor[HTML]{E3CEEF}0.700     & 0.507    & \cellcolor[HTML]{E3CEEF}0.727  & 0.500  & \cellcolor[HTML]{E3CEEF}0.673 \\
                                   & R@50$\uparrow$                & \cellcolor[HTML]{CFD3DF}0.820 & 0.663          & 0.297              & 0.633      & \cellcolor[HTML]{E3CEEF}0.773    & 0.667      & \cellcolor[HTML]{E3CEEF}0.777     & 0.620    & \cellcolor[HTML]{E3CEEF}0.807  & 0.607  & \cellcolor[HTML]{E3CEEF}0.800 \\
                                   & R@100$\uparrow$               & \cellcolor[HTML]{CFD3DF}0.863 & 0.710          & 0.357              & 0.720      & \cellcolor[HTML]{E3CEEF}0.830    & 0.737      & \cellcolor[HTML]{E3CEEF}0.867     & 0.720    & \cellcolor[HTML]{E3CEEF}0.860  & 0.667  & \cellcolor[HTML]{E3CEEF}0.857 \\
                                   & R@200$\uparrow$               & \cellcolor[HTML]{CFD3DF}0.897 & 0.780          & 0.417              & 0.807      & \cellcolor[HTML]{E3CEEF}0.900    & 0.810      & \cellcolor[HTML]{E3CEEF}0.897     & 0.780    & \cellcolor[HTML]{E3CEEF}0.887  & 0.760  & \cellcolor[HTML]{E3CEEF}0.883 \\
\multirow{-6}{*}{\textbf{SciFact}} & R@300$\uparrow$               & \cellcolor[HTML]{CFD3DF}0.913 & 0.813          & 0.457              & 0.840      & \cellcolor[HTML]{E3CEEF}0.930    & 0.857      & \cellcolor[HTML]{E3CEEF}0.920     & 0.807    & \cellcolor[HTML]{E3CEEF}0.910  & 0.810  & \cellcolor[HTML]{E3CEEF}0.913 \\ \hline
                                   & R@5$\uparrow$                 & \cellcolor[HTML]{CFD3DF}0.381 & 0.189          & 0.064              & 0.216      & \cellcolor[HTML]{E3CEEF}0.320    & 0.212      & \cellcolor[HTML]{E3CEEF}0.314     & 0.162    & \cellcolor[HTML]{E3CEEF}0.321  & 0.153  & \cellcolor[HTML]{E3CEEF}0.311 \\
                                   & R@20$\uparrow$                & \cellcolor[HTML]{CFD3DF}0.553 & 0.320          & 0.121              & 0.405      & \cellcolor[HTML]{E3CEEF}0.532    & 0.389      & \cellcolor[HTML]{E3CEEF}0.518     & 0.330    & \cellcolor[HTML]{E3CEEF}0.522  & 0.321  & \cellcolor[HTML]{E3CEEF}0.482 \\
                                   & R@50$\uparrow$                & \cellcolor[HTML]{CFD3DF}0.664 & 0.440          & 0.204              & 0.551      & \cellcolor[HTML]{E3CEEF}0.651    & 0.530      & \cellcolor[HTML]{E3CEEF}0.658     & 0.480    & \cellcolor[HTML]{E3CEEF}0.665  & 0.444  & \cellcolor[HTML]{E3CEEF}0.614 \\
                                   & R@100$\uparrow$               & \cellcolor[HTML]{CFD3DF}0.753 & 0.553          & 0.298              & 0.655      & \cellcolor[HTML]{E3CEEF}0.742    & 0.666      & \cellcolor[HTML]{E3CEEF}0.743     & 0.605    & \cellcolor[HTML]{E3CEEF}0.750  & 0.582  & \cellcolor[HTML]{E3CEEF}0.709 \\
                                   & R@200$\uparrow$               & \cellcolor[HTML]{CFD3DF}0.838 & 0.662          & 0.420              & 0.771      & \cellcolor[HTML]{E3CEEF}0.816    & 0.784      & \cellcolor[HTML]{E3CEEF}0.826     & 0.725    & \cellcolor[HTML]{E3CEEF}0.839  & 0.698  & \cellcolor[HTML]{E3CEEF}0.811 \\
\multirow{-6}{*}{\textbf{SCIDOCS}} & R@300$\uparrow$               & \cellcolor[HTML]{CFD3DF}0.873 & 0.734          & 0.518              & 0.827      & \cellcolor[HTML]{E3CEEF}0.867    & 0.844      & \cellcolor[HTML]{E3CEEF}0.870     & 0.808    & \cellcolor[HTML]{E3CEEF}0.881  & 0.775  & \cellcolor[HTML]{E3CEEF}0.861 \\ \hline
                                   & R@5$\uparrow$                 & \cellcolor[HTML]{CFD3DF}0.573 & 0.238          & 0.204              & 0.341      & \cellcolor[HTML]{E3CEEF}0.452    & 0.356      & \cellcolor[HTML]{E3CEEF}0.415     & 0.386    & \cellcolor[HTML]{E3CEEF}0.495  & 0.352  & \cellcolor[HTML]{E3CEEF}0.484 \\
                                   & R@20$\uparrow$                & \cellcolor[HTML]{CFD3DF}0.893 & 0.513          & 0.383              & 0.679      & \cellcolor[HTML]{E3CEEF}0.823    & 0.691      & \cellcolor[HTML]{E3CEEF}0.799     & 0.701    & \cellcolor[HTML]{E3CEEF}0.838  & 0.648  & \cellcolor[HTML]{E3CEEF}0.824 \\
                                   & R@50$\uparrow$                & \cellcolor[HTML]{CFD3DF}0.948 & 0.660          & 0.508              & 0.822      & \cellcolor[HTML]{E3CEEF}0.920    & 0.824      & \cellcolor[HTML]{E3CEEF}0.911     & 0.841    & \cellcolor[HTML]{E3CEEF}0.929  & 0.804  & \cellcolor[HTML]{E3CEEF}0.918 \\
                                   & R@100$\uparrow$               & \cellcolor[HTML]{CFD3DF}0.973 & 0.751          & 0.598              & 0.908      & \cellcolor[HTML]{E3CEEF}0.962    & 0.910      & \cellcolor[HTML]{E3CEEF}0.955     & 0.904    & \cellcolor[HTML]{E3CEEF}0.967  & 0.871  & \cellcolor[HTML]{E3CEEF}0.958 \\
                                   & R@200$\uparrow$               & \cellcolor[HTML]{CFD3DF}0.987 & 0.833          & 0.676              & 0.947      & \cellcolor[HTML]{E3CEEF}0.978    & 0.952      & \cellcolor[HTML]{E3CEEF}0.982     & 0.957    & \cellcolor[HTML]{E3CEEF}0.982  & 0.918  & \cellcolor[HTML]{E3CEEF}0.972 \\
\multirow{-6}{*}{\textbf{ArguAna}} & R@300$\uparrow$               & \cellcolor[HTML]{CFD3DF}0.992 & 0.875          & 0.731              & 0.961      & \cellcolor[HTML]{E3CEEF}0.987    & 0.967      & \cellcolor[HTML]{E3CEEF}0.984     & 0.974    & \cellcolor[HTML]{E3CEEF}0.985  & 0.945  & \cellcolor[HTML]{E3CEEF}0.982 \\ \hline
                                   & R@5$\uparrow$                 & \cellcolor[HTML]{CFD3DF}0.956 & 0.782          & 0.342              & 0.657      & \cellcolor[HTML]{E3CEEF}0.873    & 0.685      & \cellcolor[HTML]{E3CEEF}0.883     & 0.740    & \cellcolor[HTML]{E3CEEF}0.894  & 0.643  & \cellcolor[HTML]{E3CEEF}0.877 \\
                                   & R@20$\uparrow$                & \cellcolor[HTML]{CFD3DF}0.989 & 0.876          & 0.445              & 0.806      & \cellcolor[HTML]{E3CEEF}0.954    & 0.835      & \cellcolor[HTML]{E3CEEF}0.957     & 0.873    & \cellcolor[HTML]{E3CEEF}0.967  & 0.789  & \cellcolor[HTML]{E3CEEF}0.954 \\
                                   & R@50$\uparrow$                & \cellcolor[HTML]{CFD3DF}0.995 & 0.914          & 0.511              & 0.876      & \cellcolor[HTML]{E3CEEF}0.978    & 0.904      & \cellcolor[HTML]{E3CEEF}0.981     & 0.927    & \cellcolor[HTML]{E3CEEF}0.985  & 0.859  & \cellcolor[HTML]{E3CEEF}0.978 \\
                                   & R@100$\uparrow$               & \cellcolor[HTML]{CFD3DF}0.998 & 0.939          & 0.565              & 0.920      & \cellcolor[HTML]{E3CEEF}0.989    & 0.939      & \cellcolor[HTML]{E3CEEF}0.989     & 0.957    & \cellcolor[HTML]{E3CEEF}0.992  & 0.901  & \cellcolor[HTML]{E3CEEF}0.988 \\
                                   & R@200$\uparrow$               & \cellcolor[HTML]{CFD3DF}0.999 & 0.954          & 0.616              & 0.950      & \cellcolor[HTML]{E3CEEF}0.995    & 0.963      & \cellcolor[HTML]{E3CEEF}0.996     & 0.972    & \cellcolor[HTML]{E3CEEF}0.996  & 0.932  & \cellcolor[HTML]{E3CEEF}0.994 \\
\multirow{-6}{*}{\textbf{Quora}}   & R@300$\uparrow$               & \cellcolor[HTML]{CFD3DF}1.000 & 0.963          & 0.647              & 0.961      & \cellcolor[HTML]{E3CEEF}0.996    & 0.973      & \cellcolor[HTML]{E3CEEF}0.998     & 0.981    & \cellcolor[HTML]{E3CEEF}0.997  & 0.947  & \cellcolor[HTML]{E3CEEF}0.996 \\ \hline
                                   & R@5$\uparrow$                 & \cellcolor[HTML]{CFD3DF}0.628 & 0.361          & 0.159              & 0.235      & \cellcolor[HTML]{E3CEEF}0.530    & 0.249      & \cellcolor[HTML]{E3CEEF}0.540     & 0.203    & \cellcolor[HTML]{E3CEEF}0.512  & 0.247  & \cellcolor[HTML]{E3CEEF}0.538 \\
                                   & R@20$\uparrow$                & \cellcolor[HTML]{CFD3DF}0.806 & 0.495          & 0.225              & 0.391      & \cellcolor[HTML]{E3CEEF}0.714    & 0.406      & \cellcolor[HTML]{E3CEEF}0.716     & 0.338    & \cellcolor[HTML]{E3CEEF}0.695  & 0.389  & \cellcolor[HTML]{E3CEEF}0.707 \\
                                   & R@50$\uparrow$                & \cellcolor[HTML]{CFD3DF}0.871 & 0.584          & 0.276              & 0.497      & \cellcolor[HTML]{E3CEEF}0.801    & 0.513      & \cellcolor[HTML]{E3CEEF}0.804     & 0.438    & \cellcolor[HTML]{E3CEEF}0.778  & 0.491  & \cellcolor[HTML]{E3CEEF}0.793 \\
                                   & R@100$\uparrow$               & \cellcolor[HTML]{CFD3DF}0.907 & 0.643          & 0.316              & 0.572      & \cellcolor[HTML]{E3CEEF}0.848    & 0.581      & \cellcolor[HTML]{E3CEEF}0.853     & 0.510    & \cellcolor[HTML]{E3CEEF}0.833  & 0.573  & \cellcolor[HTML]{E3CEEF}0.840 \\
                                   & R@200$\uparrow$               & \cellcolor[HTML]{CFD3DF}0.936 & 0.699          & 0.360              & 0.661      & \cellcolor[HTML]{E3CEEF}0.886    & 0.665      & \cellcolor[HTML]{E3CEEF}0.892     & 0.592    & \cellcolor[HTML]{E3CEEF}0.874  & 0.648  & \cellcolor[HTML]{E3CEEF}0.884 \\
\multirow{-6}{*}{\textbf{NQ}}      & R@300$\uparrow$               & \cellcolor[HTML]{CFD3DF}0.949 & 0.723          & 0.383              & 0.705      & \cellcolor[HTML]{E3CEEF}0.906    & 0.708      & \cellcolor[HTML]{E3CEEF}0.909     & 0.643    & \cellcolor[HTML]{E3CEEF}0.895  & 0.701  & \cellcolor[HTML]{E3CEEF}0.904 \\ \hline
\end{tabular}%
}
\label{tab:main_exp}
\end{table*}

\subsubsection{Security}

We adopt Vec2Text \citep{morris2023text} to perform inversion attacks. 
 Table \ref{tab:exp_Inversion} shows the reconstruction quality of embeddings after being processed by our method. The results demonstrate that STEER significantly reduces reconstruction quality, with Rouge-L dropping to about $16\%$-$30\%$ of the original and BLEU scores down to about $0.8\%$-$4\%$. As model capacity increases, we observe a rise in cosine similarity and a corresponding improvement in reconstruction quality.  However, \textbf{a stronger fitting capacity does not imply a higher attack risk. Even under the best-fitting scenario, all inversion metrics remain substantially lower than the original, with only about a $\mathbf{1\%}$ drop in retrieval accuracy (Recall@300), effectively safeguarding the security of both queries and embeddings.}

\begin{table*}[h]\scriptsize
\centering
\caption{Inversion Performance in Rouge-L, BLEU and Cosine Similarity.}
\begin{tabular}{lcccccccccc}
\midrule
\textbf{Methods$\rightarrow$} & \textbf{gtr-t5-base}                   & \textbf{SnD}                   & \multicolumn{4}{c}{\textbf{all-MiniLM-L6-v2}}             & \multicolumn{4}{c}{\textbf{all-MiniLM-L12-v2}}            \\ \cmidrule(l){2-2} \cmidrule(l){3-3} \cmidrule(l){4-7} \cmidrule(l){8-11} 
\textbf{Metrics}              & \textbackslash                         & \textbackslash                 & linear & small  & medium & base                           & linear & small  & medium & base                           \\ \midrule
\textbf{Rouge-L}              & \cellcolor[HTML]{CFD3DF}77.438         & \cellcolor[HTML]{DAE8FC}24.104 & 18.598  & 19.716 & 20.127 & \cellcolor[HTML]{E3CEEF}22.085 & 17.587 & 20.426 & 20.234 & \cellcolor[HTML]{E3CEEF}21.982 \\
\textbf{BLEU-1}               & \cellcolor[HTML]{CFD3DF}74.170         & \cellcolor[HTML]{DAE8FC}25.590 &16.750  & 17.570 & 17.230 & \cellcolor[HTML]{E3CEEF}17.700 & 15.590 & 16.590 & 16.170 & \cellcolor[HTML]{E3CEEF}16.900 \\
\textbf{BLEU-2}               & \cellcolor[HTML]{CFD3DF}55.700         & \cellcolor[HTML]{DAE8FC}5.160  & 2.690  & 3.300  & 2.990  & \cellcolor[HTML]{E3CEEF}3.960  & 2.360  & 2.480  & 2.980  & \cellcolor[HTML]{E3CEEF}3.890  \\
\textbf{BLEU}                 & \cellcolor[HTML]{CFD3DF}52.490         & \cellcolor[HTML]{DAE8FC}1.980  & 1.200  & 1.600  & 1.600  & \cellcolor[HTML]{E3CEEF}2.560  & 1.070  & 0.780  & 1.500  & \cellcolor[HTML]{E3CEEF}2.210  \\
\textbf{cos}                  & \cellcolor[HTML]{CFD3DF}\textbackslash & \cellcolor[HTML]{DAE8FC}0.526  & 0.531 & 0.575  & 0.617  & \cellcolor[HTML]{E3CEEF}0.684  & 0.510  & 0.575  & 0.610  & \cellcolor[HTML]{E3CEEF}0.678  \\ \hline
\midrule
\textbf{Methods$\rightarrow$} & \textbf{gtr-t5-base}                   & \textbf{SnD}                   & \multicolumn{4}{c}{\textbf{bge-small-en}}                 & \multicolumn{4}{c}{\textbf{e5-small}}                     \\  \cmidrule(l){2-2} \cmidrule(l){3-3} \cmidrule(l){4-7} \cmidrule(l){8-11} 
\textbf{Metrics}              & \textbackslash                         & \textbackslash                 & linear & small  & medium & base                           & linear & small  & medium & base                           \\ \midrule
\textbf{Rouge-L}              & \cellcolor[HTML]{CFD3DF}77.438         & \cellcolor[HTML]{DAE8FC}24.104 & 12.757 & 19.466 & 20.117 & \cellcolor[HTML]{E3CEEF}23.694 & 17.559 & 20.715 & 20.597 & \cellcolor[HTML]{E3CEEF}23.583 \\
\textbf{BLEU-1}               & \cellcolor[HTML]{CFD3DF}74.170         & \cellcolor[HTML]{DAE8FC}25.590 & 15.180 & 22.390 & 15.810 & \cellcolor[HTML]{E3CEEF}18.780 & 16.940 & 18.390 & 17.290 & \cellcolor[HTML]{E3CEEF}18.150 \\
\textbf{BLEU-2}               & \cellcolor[HTML]{CFD3DF}55.700         & \cellcolor[HTML]{DAE8FC}5.160  & 1.490  & 3.590  & 3.180  & \cellcolor[HTML]{E3CEEF}4.200  & 2.910  & 3.590  & 3.040  & \cellcolor[HTML]{E3CEEF}4.050  \\
\textbf{BLEU}                 & \cellcolor[HTML]{CFD3DF}52.490         & \cellcolor[HTML]{DAE8FC}1.980  & 0.450  & 1.300  & 1.570  & \cellcolor[HTML]{E3CEEF}2.670  & 1.710  & 2.160  & 1.560  & \cellcolor[HTML]{E3CEEF}2.430  \\
\textbf{cos}                  & \cellcolor[HTML]{CFD3DF}\textbackslash & \cellcolor[HTML]{DAE8FC}0.526  & 0.396  & 0.517  & 0.580  & \cellcolor[HTML]{E3CEEF}0.680  & 0.476  & 0.577  & 0.616  & \cellcolor[HTML]{E3CEEF}0.676  \\ \hline
\end{tabular}
\label{tab:exp_Inversion}
\end{table*}

\subsection{Ablation Study}

We conducted three main ablation studies under the STEER framework to further validate its effectiveness, using the low-dimensional embedding model all-MiniLM-L12-v2.

\begin{figure}[h]
    \centering

     \begin{subfigure}[b]{0.35\columnwidth}
        \centering
        \includegraphics[width=\textwidth]{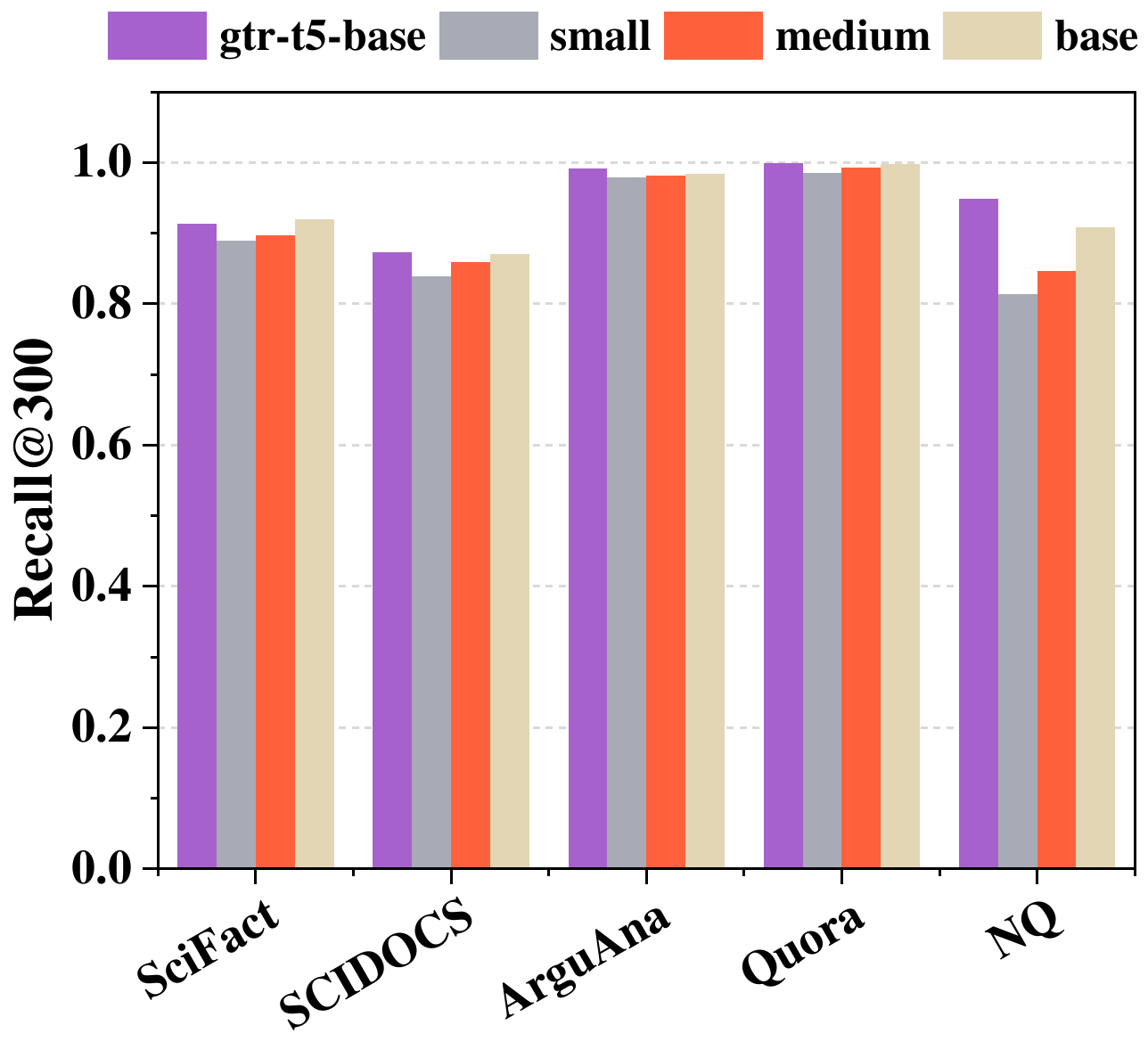}
        \caption{Retrieval Performance}
        \label{fig:retrieval_mlp}
    \end{subfigure}
    \hspace{0.01\columnwidth} 
    \begin{subfigure}[b]{0.35\columnwidth}
        \centering
        \includegraphics[width=\textwidth]{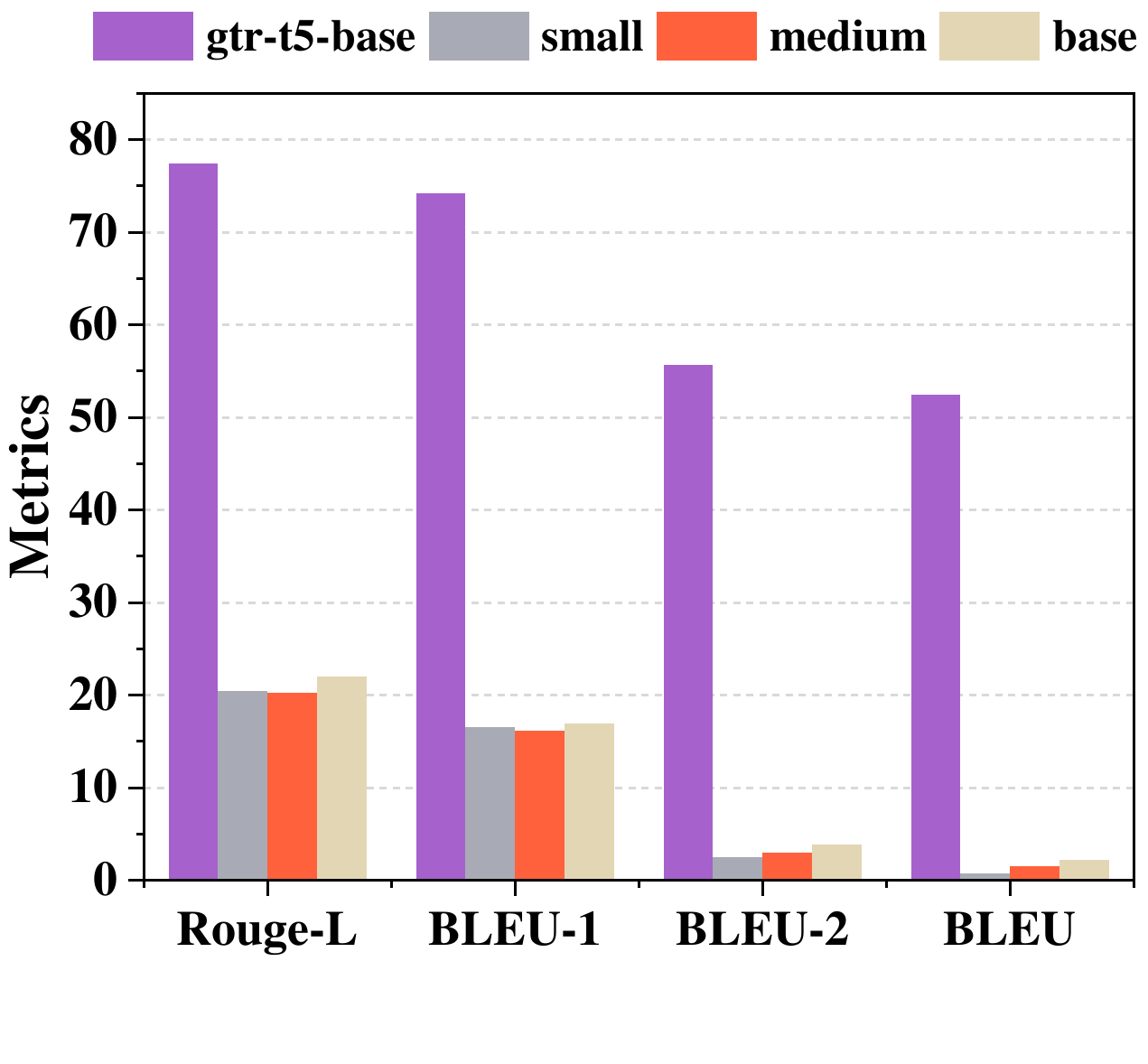}
        \caption{Inversion Performance}
        \label{fig:inversion_mlp}
    \end{subfigure}
      \caption{The Performance of Nonlinear Models with Varying Fitting Capacities.
}
    \label{fig:Varying_Fitting_Cap}
\end{figure}

\textbf{STEER Using Different MLPs.} Figure \ref{fig:Varying_Fitting_Cap}  shows that as the model’s fitting capacity increases, retrieval performance consistently improves and\textbf{ remains comparable to that of gtr-t5-base across all datasets. } This indicates that our structured semantic alignment mapping can accurately reconstruct the geometry of the target embedding space, maintaining retrieval accuracy and demonstrating strong generalization. Figure \ref{fig:retrieval_mlp} illustrates that even as the fitting ability improves, Rouge-L and BLEU remain significantly lower than directly exposing raw embeddings. This suggests that our embedding perturbation is highly structured and effectively mitigates the risk of EIAs. \textbf{These trends are consistent across models and datasets, confirming that STEER enhances security while preserving robustness and generalizability across different deployment settings.}

\textbf{Comparison between Different Anonymization Levels and STEER.} We levearaged three different anonymization levels (Level 1/2/3), with Level 1 being the strictest; details are provided in  Appendix \ref{sec:desensitization} . Figure \ref{fig:desensitize} shows that anonymization severely degrades retrieval accuracy by removing key query information. For example, on SCIDOCS and NQ, Recall@100 under Level 1 drops below $0.4$—far below our method. Even Level 3 fails to restore acceptable performance. \textbf{This highlights that while anonymization avoids exposing raw text, it causes irreversible semantic loss. In contrast, STEER preserves semantic content without revealing the original text, maintaining high retrieval performance across multiple datasets.}

\begin{figure}[h]
    \centering

     \begin{subfigure}[b]{0.35\columnwidth}
        \centering
        \includegraphics[width=\textwidth]{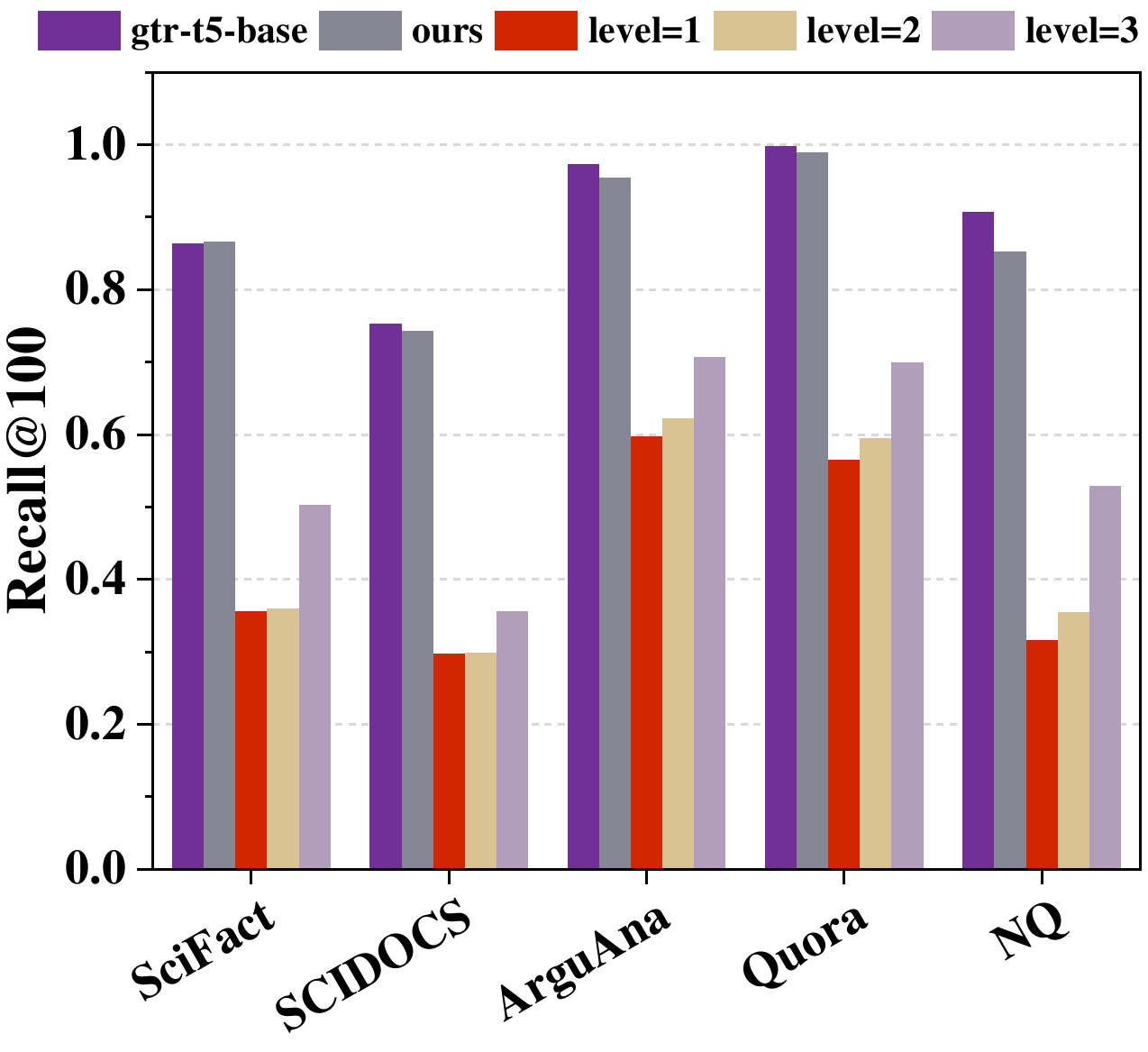}
        \caption{Recall@100}
        \label{fig:recall100}
    \end{subfigure}
    \hspace{0.01\columnwidth} 
    \begin{subfigure}[b]{0.35\columnwidth}
        \centering
        \includegraphics[width=\textwidth]{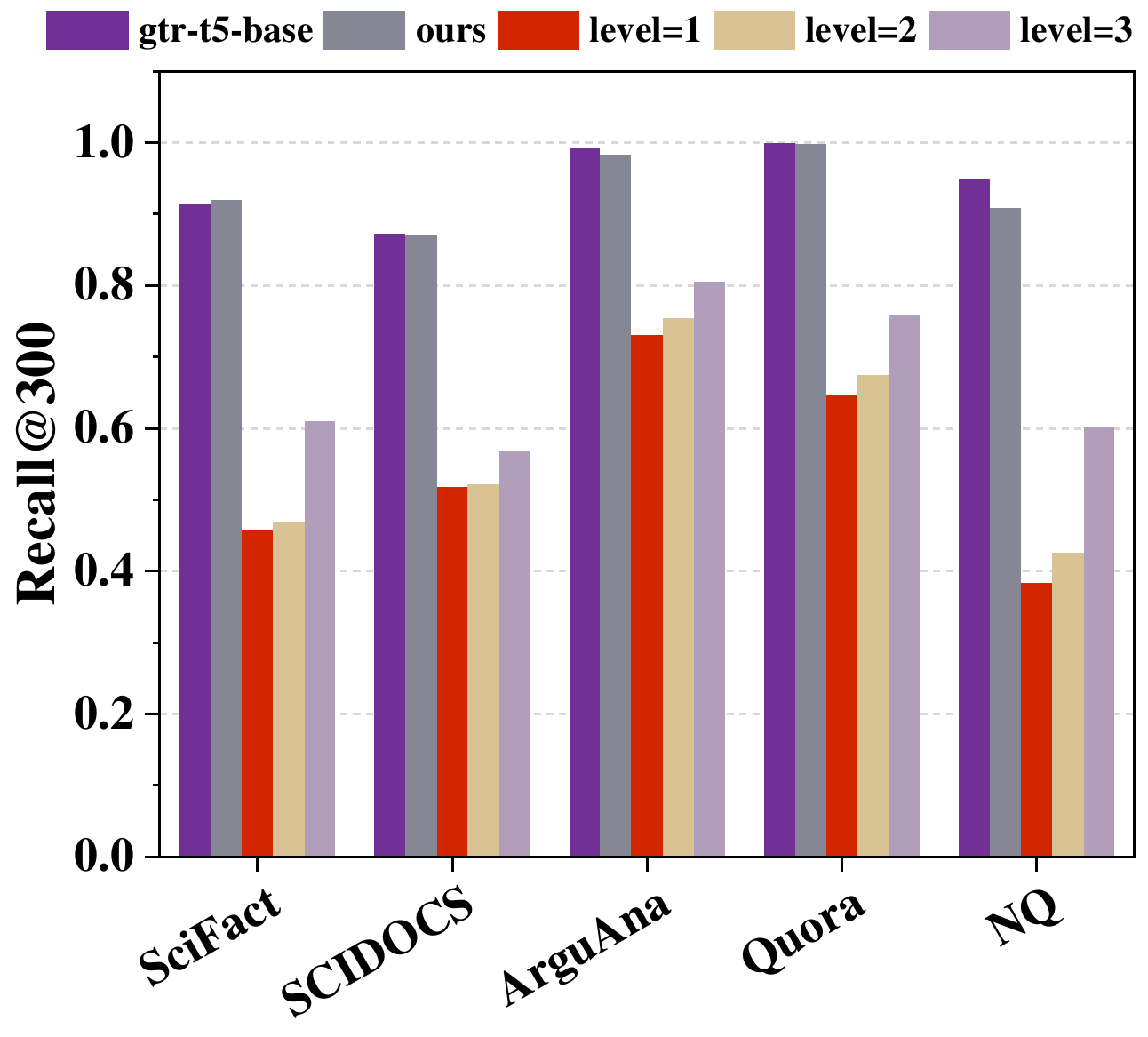}
        \caption{Recall@300}
        \label{fig:recall300}
    \end{subfigure}
      \caption{The Retrieval Performance under Varying Levels of Desensitization.}
    \label{fig:desensitize}
\end{figure}

\begin{figure}[h]
    \centering

     \begin{subfigure}[b]{0.35\columnwidth}
        \centering
        \includegraphics[width=\textwidth]{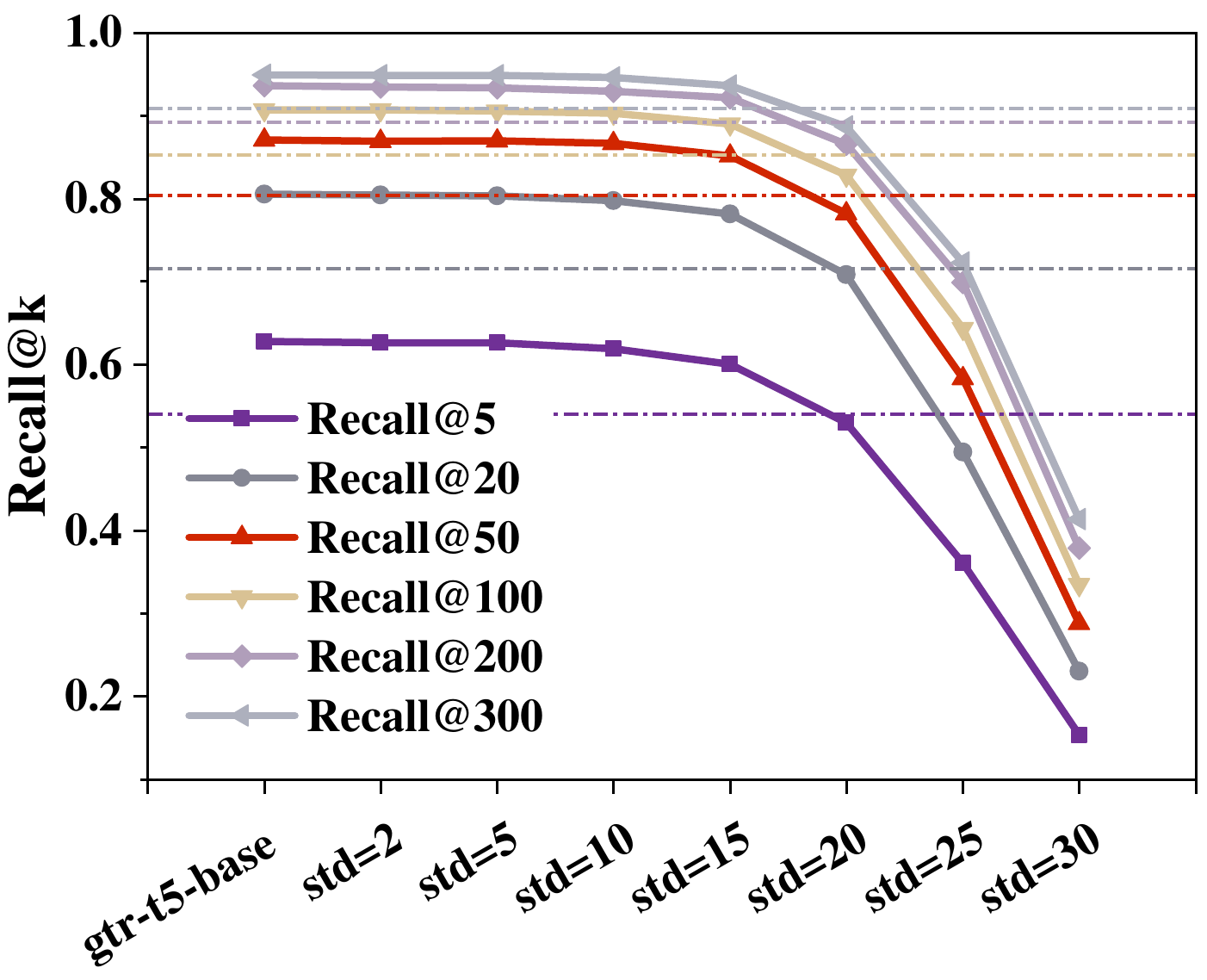}
        \caption{Retrieval Performance}
        \label{fig:noise_retrieval}
    \end{subfigure}
    \hspace{0.01\columnwidth} 
    \begin{subfigure}[b]{0.35\columnwidth}
        \centering
        \includegraphics[width=\textwidth]{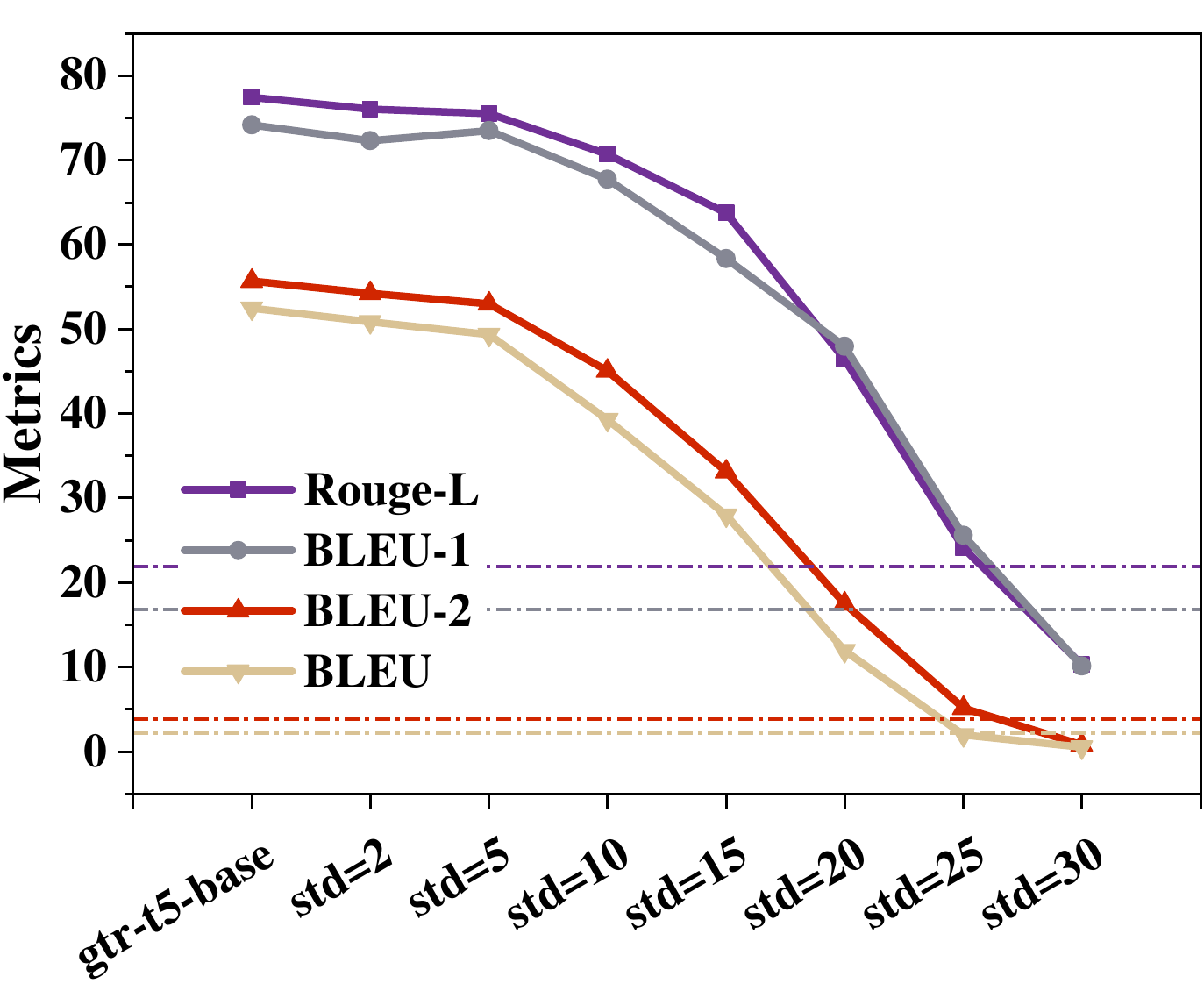}
        \caption{Inversion Performance}
        \label{fig:noise_inversion}
    \end{subfigure}
      \caption{SnD under Varying Noise Levels.}
    \label{fig:Snd_noise}
\end{figure}

\textbf{Comparison between Different SnD Noise Levels and STEER.} In the SnD scheme, it is crucial to select the noise strength to balance privacy and utility. Thus, we adopted different noise levels and measured both retrieval accuracy and inversion risk. As shown in Figure \ref{fig:Snd_noise}, stronger noise leads to lower BLEU and Rouge-L scores, indicating reduced inversion risk, but also results in a sharp decline in Recall@k. In contrast, our STEER (dashed lines) achieves significantly better retrieval performance under comparable inversion risk levels. This suggests that, \textbf{compared to the unstructured perturbations used in SnD, STEER provides more stable and effective privacy protection without the need for random noise, while still preserving retrieval accuracy}.

\section{Discussion}
\label{sec:discussion}

Based on the threat model defined in our work, text embeddings stored in VDB are still generated by the embedding model of the VDB service provider. This embedding model used for retrieval remains under the service provider's control, while users only employ local models to generate approximate embeddings to mitigate EIAs. Naturally, this raises a question: \textit{If users are capable of deploying the embedding model locally to generate both the embeddings of VDB and query embeddings, would security risks persist?} Recent studies proposed by \citet{jha2025harnessing} and \citet{zhang2025universal} solve this question. Their embedding inversion methods can still infer and extract text information even if the embedding model is deployed locally. We believe that even in such a new threat model, STEER can still resist EIAs to a certain extent. This is because all embeddings obtained by the attacker are approximate embeddings. The approximate embedding space does not belong to any existing embedding model, making it difficult for attackers to reconstruct text information.

\section{Conclusion}
\label{sec:conclusion}

In this paper, we propose STEER, a novel and privacy-preserving vector retrieval framework designed to protect user query text privacy. During the $\mathsf{setup}$ $\mathsf{phase}$, a local embedding model is deployed on the user’s device, and a set of non-sensitive, general-purpose texts is used to generate embedding pairs from both the local and server-side models. In the $\mathsf{transformation}$ $\mathsf{phase}$, the mapping function between embedding spaces is computed locally using the embedding pairs prepared in the $\mathsf{setup}$ $\mathsf{phase}$. When a query is issued, STEER first encodes it using the local model, and then applies the learned transformation to generate an approximate embedding in server's embedding space.  The retrieval is finally performed within the original VDB using this approximate embedding. Extensive experiments demonstrate that STEER has no need to transmit query text to the server while resisting EIAs. Crucially, it preserves the semantic integrity of the embeddings to maintain retrieval performance. Our work demonstrates that the deviation introduced by transformation defends against EIAs while significantly improving the retrieval accuracy over current methods.

\bibliography{iclr2025_conference}
\bibliographystyle{iclr2025_conference}

\appendix

\section*{Appendix}

\section{ Definition of Desensitization Levels}\label{sec:desensitization}

To evaluate the impact of privacy-preserving preprocessing, we adopt a rule-based desensitization framework that categorizes identifiable spans into three semantic sensitivity levels: High, Medium, and Low. The classification is based on named entity types and pattern-based matching. The anonymization level determines the extent of perturbation applied to the input query.

\textbf{Level 1 (Strict).}
All entities labeled as High, Medium, or Low sensitivity are anonymized. This includes personally identifiable information (PII) such as email addresses, phone numbers, ID numbers, IP addresses, and financial tokens (High), names, organizations, locations, and online handles (Medium), as well as generic numeric or nominal expressions (Low).

\textbf{Level 2 (Moderate).}
Only Medium and Low sensitivity entities are processed. Highly sensitive entities such as email addresses or identity numbers are retained to preserve more semantic content while still reducing surface identifiability.

\textbf{Level 3 (Light).}
Only Low sensitivity tokens, such as dates, quantities, and common nouns are anonymized. This minimally alters the semantic structure and serves as a lower bound in the privacy–utility trade-off.

Each sensitivity level is handled using a distinct anonymization operator. High-sensitivity spans are replaced with placeholders. Medium-sensitivity spans are partially masked. Low-sensitivity spans are replaced with generic semantic tokens. The desensitization pipeline integrates SpaCy-based named entity recognition and regular expression matching and is applied line-by-line to all input queries during evaluation.

\section{Model Size of Mapping Functions}\label{sec:model_size}

To study how differences in the fitting capacity of the mapping function affect retrieval performance and security robustness, we design four alignment models with varying capacities. The model sizes are shown in Table
\ref{tab:Model Size}.

\begin{table}[h]
\centering
\begin{tabular}{@{}ccc@{}}
\toprule
\multicolumn{2}{c}{\textbf{Model}}          & \textbf{Size} \\ \midrule
Linear                     & \textbackslash & 2.3MB         \\ \midrule
\multirow{3}{*}{Nonlinear} & Small          & 5.7MB         \\
                           & Medium         & 12.33MB       \\
                           & Base           & 105.51MB      \\ \bottomrule
\end{tabular}
\caption{Model Size.}
\label{tab:Model Size}
\end{table}

\section{Case Study: Examples of EIAs }

Examples of EIAs under nonlinear setting (base) with all-MiniLM-L12-v2 are as follows:

\begin{tcolorbox}[colframe=blue!50!black, colback=blue!10!white, coltitle=white, sharp corners]
{ \textbf{Original:}  what happened to regina's mother on switched at birth}

\textbf{Inversion: }when she drank on her, she woke up to find that she was going to be dead when she was told she was going to switch mom
\end{tcolorbox}

\begin{tcolorbox}[colframe=blue!50!black, colback=blue!10!white, coltitle=white, sharp corners]
{ \textbf{Original:}  where is fe best absorbed in the body

\textbf{Inversion: }body the GF concentration allows the absorption of iron at a favourable level. Citing the literature on the surface of the skin as a
}
\end{tcolorbox}

\begin{tcolorbox}[colframe=blue!50!black, colback=blue!10!white, coltitle=white, sharp corners]
{ \textbf{Original:}  is kermit the frog part of sesame street

\textbf{Inversion: }Chuck Bruno character Beet Street. Sesame plays a minor part in the a.k.a. 'Kung Fu Panda
}
\end{tcolorbox}

\end{document}